*Article*

# Shape Sensing for Continuum Robotics using Optoelectronic Sensors with Convex Reflectors

Dalia Osman [1], Xinli Du [1], Timothy Minton [1] and Yohan Noh [1,*]

[1] Department of Mechanical and Aerospace Engineering, Brunel University London, UK
* Correspondence: yohan.noh@brunel.ac.uk

**Abstract:** Three-dimensional shape sensing in soft and continuum robotics is a crucial aspect for stable actuation and control in fields such as Minimally Invasive surgery, as the estimation of complex curvatures while using continuum robotic tools is required to manipulate through fragile paths. This challenge has been addressed using a range of different sensing techniques, for example, Fibre Bragg grating (FBG) technology, inertial measurement unit (IMU) sensor networks or stretch sensors. Previously, an optics-based method, using optoelectronic sensors was explored, offering a simple and cost-effective solution for shape sensing in a flexible tendon-actuated manipulator in two orientations. This was based on proximity-modulated angle estimation and has been the basis for the shape-sensing method addressed in this paper. The improved and miniaturized technique demonstrated in this paper is based on the use of a spherically shaped reflector with optoelectronic sensors integrated into a tendon actuated robotic manipulator. Upgraded sensing capability is achieved using optimization of the spherical reflector shape in terms of sensor range and resolution, and improved calibration is achieved through the integration of spherical bearings for friction-free motion. Shape estimation is achieved in two orientations upon calibration of sensors, with a maximum Root Mean Square Error (RMS) of 3.37°.

**Keywords:** Shape Sensing; Optoelectronic Sensors; Continuum Robotics

## 1. Introduction

A multitude of shape sensing systems have been developed for continuum and hyper redundant robots. For safe and accurate utilization of a continuum robot, a stable position control system must be integrated. As the distinguishing features of continuum and hyper-redundant robots are narrow dimensions, flexibility, and use of soft, compliant materials, the integration with a position shape sensing system that does not compromise these features is challenging. Continuum robots have been utilised in difficult environments, such as search and rescue missions, as well as for exploration missions in unknown environments [1][2], and in maintenance for onsite inspection and repairs of large machines such as aeroplane turbine engines, which are difficult to access internally with rigid tools due to their confined environment and may otherwise be time-consuming and costly to dismantle [3]. A popular field in which continuum robots are used is for minimally invasive surgery (MIS) [4]. In MIS, a small incision is made in the body through which tools are passed for carrying out surgical procedures. Typically, shape sensing systems for MIS based continuum robots must be miniaturized in order to allow full motion and flexibility capabilities through complex paths, and use of actuators and encoders directly within joints along the robotic manipulator as is done in larger rigid robots is not feasible in such applications.

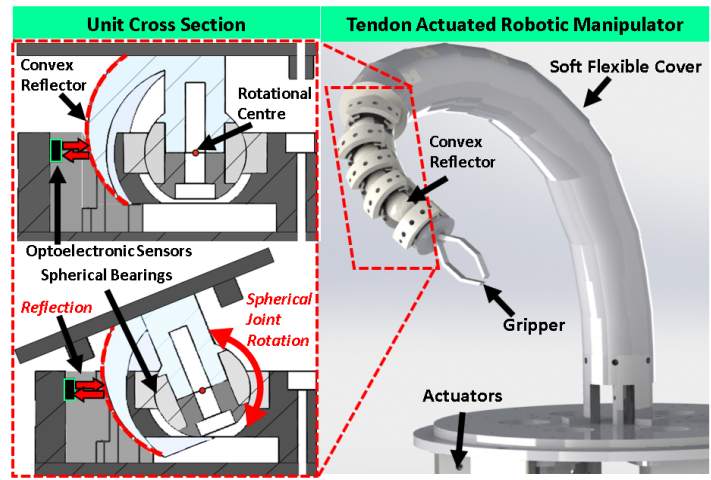

**Fig.1.** Optoelectronic shape sensing for a tendon actuated robot. A spherically convex shaped reflector is shown along with the optoelectronic sensor placement. (Designed in *Onshape*)

A variety of sensor types have been employed in the design of shape sensing systems. Microelectromechanical (MEMs) tracking sensors such as inertial sensors comprise of a system of accelerometers, gyroscope, and magnetometers to measure multiaxial orientation measurements. Some examples demonstrate a shape sensing approach for hyper redundant, snake-like, and pneumatic robots using a network of IMU sensors fixed along the links of the robot, using the sensor information to calculate the tip pose through a kinematic model [5][6][7]. These sensor sizes are relatively small, although fusion with fully soft manipulators may be more difficult. Inertial sensors have also been used in SensorTape [8], a flexible modular tape embedded with a network of sensors including inertial sensors, that can be attached to sections of a flexible robotic structure for shape sensing in three-dimensional space. Inertial sensors however can suffer from error due to magnetic interference, gyroscopic drift as well as position estimation errors from mechanical vibration [9].

Stretchable strain sensors have been employed in some soft robotic manipulators. Jinho So et al, for example used stretchable sensors on a section of a flexible manipulator for use in MIS [10]. The skin-like stretch sensor was composed of carbon nanotube layers within a silicone sheet, placed equidistantly around a silicone manipulator. These types of sensors respond to change in strain upon bending of the manipulator, as a change in resistance is measured due to the deformation. Many types of materials have been used to develop such stretch sensors, including silver nanowire, carbon black, as well as eutectic gallium-indium liquid. Although highly flexible, the inherent properties of these materials exhibit signal non-linearity, and noise due to inhomogeneities in the material. Another point is the assumption of constant curvature in the geometrical modelling of the elastic backbones of the soft robot, which means that more complex curves would not be able to be measured using such stretch sensors.

Some other innovative techniques made use of marker tracking cameras. This utilised passively attached tendons that were only attached on one end to the body of the robotic segment. The lower end of these passive tendons had markers that were detected by cameras to measure their linear displacement during bending of the robot [11]. This allowed more accurate measurement of the robot's shape and torsion compared to conventional methods of measuring the length changes of the actuating tendons during bending of the continuum robot, as the passive tendons are less affected by slack or excess tension. The paper however reported an increase in stiffness of the robot due to the number of passive tendons, and in shape estimation error due to the clearance between the tendons and routing paths. Besides, if future prototypes introduce more than one segment, the passive tendons will need to pass through multiple segments, and this may cause coupling between the markers that are detected by the camera as they will be affected by motion from other robotic segments, which could reduce shape sensing estimation accuracy.

Fibre-Bragg gratings fused within optical fibres can measure curves due to shift in measured wavelength of light upon applied strain due to bending. The fibres can be integrated into narrow robotic structures and are flexible, allowing use in robotic surgical procedures such as catheterization [12]. However, these gratings are difficult to engrave, and require a specialist for accurate placement, thereby making the manufacturing process more difficult, and its signal detection equipment so called interrogator is expensive. Also, under larger curvatures, performance of shape sensing may be reduced, and error can occur due to temperature and central wavelength drift over time [13]. FBG sensors are highly sensitive to material properties and external conditions. Shape sensing errors can arise due to large bending curvatures or low stiffness conditions. They are limited to smaller bending ranges within the elastic limit of the material, and must not exceed load or external force limitations, otherwise calibration will not be maintained. FBGs are highly sensitive to any external force or load that is applied to them. This can cause issues in certain conditions. For example, if calibration of an FBG based robotic structure is completed under free space or using known curvatures, any other load or external/tangential force condition applied to the surface of the structure which is different to the calibration conditions may cause the FBGs to not accurately measure the deformations. This would lead to shape estimation errors or noise.

Optoelectronic sensors are another optical based technique that can be used for shape sensing. Prior work by the authors introduced the integration of optoelectronic reflective sensors into a tendon actuated robot for shape estimation in three dimensions [14]. Three optoelectronic sensors were placed equidistantly around each rotating link of the manipulator. These sensors comprise of an infrared LED, and a reflection detecting coupled phototransistor. For shape sensing, the change in angle between each link was estimated through detection of light reflected between the reflective surface on each consecutive link of the manipulator. This light intensity modulation due to change in proximity between the sensors and the reflecting surface allowed estimation of two orientations, through a calibration process as described in that paper. Results of the work showed the viability of the technique for shape sensing and has inspired the new technique to be demonstrated in the following sections.

Before proceeding with the new shape sensing technique, areas of improvement upon the previous method [14] were identified. One was that three sensors were required per link of the robotic manipulator, and these were embedded directly onto the disk. In a longer length of the tendon robot, the number of optoelectronic sensors would accumulate to an excessive number of sensors and electric wires. These also occupied a lot of space on the manipulator, which limited the space for the passing of tools in future development. A second point was that the links were jointed using a 3D-printed ball joint – the components were designed and printed using PLA. This resulted in a lot of friction during motion of the prototype and affected calibration measurement due to excessive vibration. Lastly, each sensor was individually installed in each unit, resulting in excessive wiring. For these reasons, in this paper, the authors will present a new optoelectronic based shape sensing mechanism to resolve the flaws of the previous version as mentioned above.

## 2. Design of a new shape sensing mechanism

*2.1 Design Concept*

The new design proposed in this paper, as illustrated in Figure 1, uses light intensity modulation between optoelectronic sensors and a convex shaped reflector, for estimation of orientation of the continuum robot segment. The design of this shape sensing system improves on the author's previous work presented in [14]. Implementation of these improvements is with the goal of achieving the following conditions: (1) That the shape sensing system can measure motion in two degrees of freedom (2) elimination of noise due to friction during motion of the robotic segment, as well as from external interference, (3) enable miniaturization of the shape sensing mechanism, and (4) simplification of the shape sensing system. These points have been addressed in the following way, in comparison to [14].

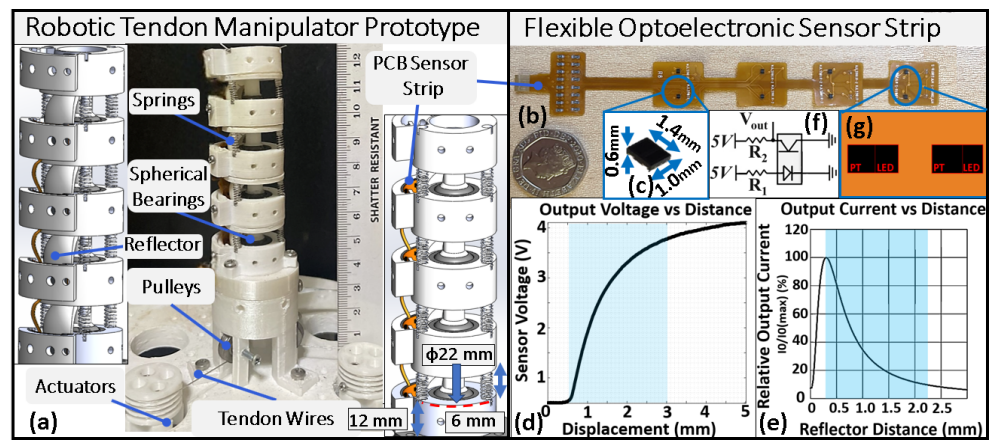

**Fig.2.** Tendon actuator robot (a) with integrated optoelectronic sensor strip (b) for shape sensing, using NJL5901R-2 (c, f) optoelectronic sensors. (d-e) show characteristic sensor output.(g) shows optoelectronic sensors positioning on PCB.

Firstly, the ball joints have been replaced by spherical bearings (GE6-DO 6mm Bore Spherical Bearing, 14mm O.D, INA). These spherical bearings are fixed within the rotational unit and linked to the consecutive unit though a shaft that fits into the spherical bore. This vastly improves motion, and friction is virtually eliminated, allowing smooth rotation between joints. More importantly, the sensing configuration in the new system is based on two sensors per rotational unit rather than three. These sensors are fitted vertically (Figure 2a), and massively reduce the occupied space on each link. All sensors were incorporated onto a single flexible circuit (Figure 2b), thereby eliminating wiring for a more simplified design. A new model of optoelectronic reflective sensor was chosen – the NJL5901R-2 (1 x 1.4 x 0.6 mm, New Japan Radio) (Figure 2c), as opposed to previously used QRE113 ON Semiconductor (3.6 x 2.9 x 1.7 mm) and is much smaller, allowing the reduction of space occupied by sensors. As described in [14], the sensing range of the previously used QRE113 optoelectronic sensor was between 2-10 mm. This large sensing range meant that during motion, there was a possibility of interference in the sensor signal from external sources, or reflection of infrared light from nearby objects or cover for enclosing the robotic structure, and this affected the sensor calibration. It is for this reason that the NJL5901R-2 optoelectronic sensor is chosen in the current prototype, as the sensing range is much shorter (0.5-3 mm) (Figure 2d) and eliminates the risk of interference from unintended reflection or scattering of light from nearby objects other than the reflector, and further enables miniaturization. Due to the housing of the sensors within the structure of the continuum robot (Figure 2a), this also protects the sensor signals from interference from external infrared sources.

In keeping with the same sensing principle, that is, light intensity modulation upon change in proximity, a new surface reflector shape was designed as part of the tendon actuated robot links. This new reflector takes on a spherical shape and is essentially a convex reflector (Figure 3a). The inner part of this spherical reflector section is concentric with the centre of rotation of the rotational unit. However, the outer bound of the spherical reflector section that is facing the sensors, has the centre of its sphere slightly offset from the rotational centre as seen in Figure 3b. This has the effect of a sphere with gradually changing radius upon rotation, relative to the sensors. As such, this means that during rotation, the distances between the sensors and the reflector change, providing a means for modulation of the reflected light intensity, that can be transformed to orientation estimation in both pitch and roll rotations. The sensing principle is based on non-contact light intensity proximity sensing. This means that the sensors do not directly measure properties such as strain and are not affected by any loads or material limitations. This means that even after calibration of the sensors in free space, the calibration matrix remains intact to accurately estimate shape regardless of any external loads or forces applied to the surface of the robotic structure.

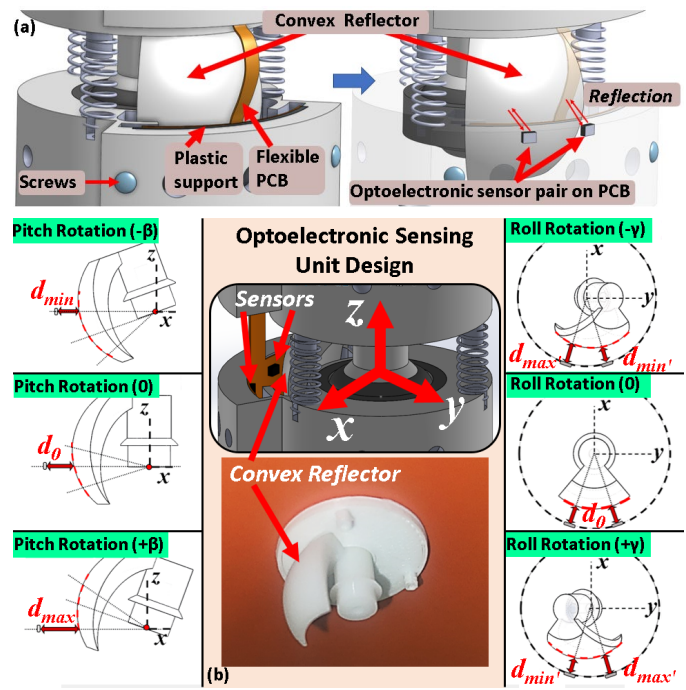

**Fig.3.** (a) Integration of optoelectronic PCB into robotic structure. (b) Cross sectional views of light path from sensor to reflector, under different orientation rotations.

This is further described in section 3. This sensing configuration along with use of the convex reflector allows use of the sensors' small proximity range (0.5 – 3mm), without compromising on the rotational range of each unit of the robotic manipulator, since considering the authors' previously used sensing configuration [14], use of a smaller sensor proximity range would require a very small distance between each consecutive rotational unit of the robotic structure, which would limit the bending range of each rotational unit.

*2.2 Shape Sensing Integration into Robotic Manipulator*

The design of the tendon actuated robot comprised of three DC motors to actuate three wire tendons. For verifying the sensing principle of the new shape sensing mechanism, four consecutive units are used as a one segment as illustrated in Figure 2a, with the fitted flexible sensor circuit shown (2b). With two sensors allocated to each disk, these total 8 sensors. The four units were linked together using the spherical bearings to form a ball joint. Each unit measured 22 mm in diameter, and 12 mm in height (Fig 2a). Each unit consisted of the spherical reflector, with a central shaft for fitting into the spherical bearing bore of the subsequent lower unit. Three tendons were routed along the structure, with a set of three springs between each unit fitted to limit some of the torsion motion. The springs between each unit were 6mm in height. Each tendon was wound around a pully system attached to the motor horn of three DC servo motors (Dynamixel XL430-W250T) at the base platform. These three tendons were used to achieve actuation in both pitch and roll orientations. A channel down one side of the structure was used to allow the optoelectronic PCB strip, shown in Figure 2, to be fitted. This was a printed flexible sensor circuit strip, comprising of four pairs of the optoelectronic sensor. The sensor strip was fixed along the channel at each unit of the robotic structure using two small screws (Figure 3a). On the top unit, a frame is fitted for allowing the IMU sensor (LPMS B, LP-RESEARCH Inc, Tokyo, Japan) to be fixed. The platform components were designed using CAD software (Onshape) and other than the spherical reflector component, were 3D printed using white PLA (Polylactic Acid) plastic. The spherical reflector component was printed using a UV Resin SLA 3D printer, with white coloured resin, for high resolution surface finish in order to maximise reflectivity and reduce noise in the sensor signal. For actuation – motion was controlled using software developed in Python, with real-time interfaced motor control, optoelectronic sensor recordings, as well as IMU sensor recordings. To achieve motion, an input target orientation of the top plate is chosen, and a vector-based model is used to calculate the required tendon lengths for conversion to three target motor positions. As the three motors move simultaneously, each pulls a tendon wire over the pulley for continuous motion.

## 3. Sensing Principle

### 3.1. Sensor Design Principle

The spherical convex reflector design is based on the work demonstrated in [15], which is on the design of a curved joint angle measurement sensor. This one-dimensional angle sensor demonstrated the potential for adapting the output of optoelectronic sensors for increased sensing sensitivity and range by altering the curvature of a reflecting surface. This concept is therefore applied here in three dimensions, where the sensor reflector is designed as a section of a sphere and can be regarded as a convex reflector. Two sets of cross sections of the reflector design are shown in Figure 3b. These illustrate the rotations in pitch (about $y$ axis) and roll (about the $x$ axis). Two optoelectronic sensors are placed circumferentially around the origin of rotation $O$, on the $xy$ plane. Light is reflected from the spherical reflector section (red dotted) (Fig 3b) and recorded by the sensors. Considering the schematic in Figure 4a, the reflector with centre $C$ is placed at a location offset to the rotation origin $O$. As such, when the reflector rotates about origin $O$, either in the pitch or roll orientations, the distance between the sensor and the reflector varies. This displacement variation with change in orientation allows change in sensor voltage readings due to varied light intensity for proximity-based sensing. In the cross sections shown in Figure 3b, the sensor is placed at an original distance of $d_0$ (0°). The rotating reflector is designed to rotate ±15° degrees about both the y axis (pitch) and x axis (roll). At these points, the reflector is at distance $d_{max}$/ $d_{min}$ and $d_{max'}$/ $d_{min'}$ from the sensor, respectively. For pitch rotation, the distances $d_0$, $d_{min}$ and $d_{max}$ correspond to the range identified in the graph in Figure 2d, highlighted in blue. This is the characteristic optoelectronic sensing graph showing sensor voltage output with increasing linear displacement from a reflector. This response is in part based on the resistors of the optoelectronics sensors – these are the LED resistor (R1), and the Phototransistor resistor (R2) (Fig 2f). These were chosen to be 680 Ω and 10 kΩ, respectively. To maximise range and sensitivity, the highlighted section (blue) of Fig 2d identifies the ideal sensor proximity to a reflector. This range identifies the ideal placement of the optoelectronic sensor with the spherical reflector. Given these distance ranges, the spherical radius and centre C of the reflector could be identified and designed. During roll motion, the proximity change in the range of ± 15 ° is $d_0$, $d_{min'}$ and $d_{max'}$, and is smaller compared to the proximity range during pitch rotation ($d_0$, $d_{min}$ and $d_{max}$).

### 3.2. Mathematical Theory

The mathematical theoretical concept behind this sensing principle is simply based on the gaussian reflection of light against a convex reflector [15][16]. Figure 4a illustrates a schematic of the sensing configuration and shows a cross sectional view of one sensor placement around the rotational spherical reflector component. It shows a system of how to calculate distance 'd' between the sensor and reflector. Given the centre of rotation O, vectors $f$, between O and the sensor, as well as vector $s_0$, between O and the spherical reflector centre C, can be known. Upon rotation of the unit about the origin in both pitch and roll orientations, a new vector $s_1$ can be calculated using Equation 1, using rotation matrices Rx and Ry. From this, distance d can be calculated as shown in Equations 2 and 3.

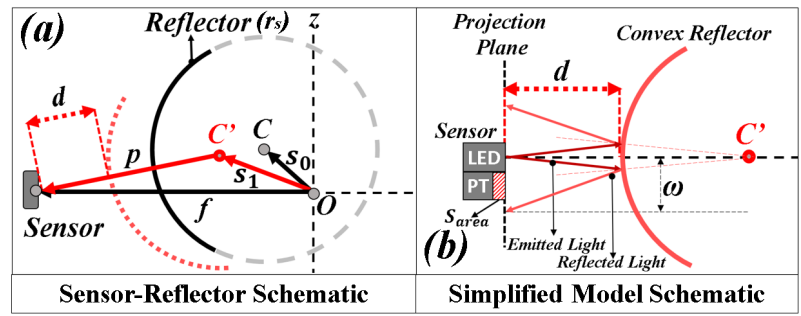

**Fig. 4.** (a) Schematic diagram for calculation of proximity 'd' between the sensor and reflector (with radius $r_s$) upon rotation.(b) Simplified schematic of light intensity model.

$$s_1 = R_x R_y s_0 \quad (1)$$

$$p = f - s_1 \quad (2)$$

$$d = |p| - r_s \quad (3)$$

Considering Figure 4b, the schematic can be simplified, by making a few assumptions, to allow the set-up of a mathematical model to describe the sensing principle. As the LED size on the sensor is quite small, we can assume that light emitted from this to be of a point source. We can also assume, due to the small scale, that the area upon which the light is incident on the reflector is small, and so it is taken to be a spherical convex reflector with a constant radius about centre C'. Using this, and values for d, as well as known phototransistor area ($s_{area}$) (Fig 4b), a theoretical reflected light intensity can be calculated using the gaussian power equation in Equation 4 [15]. This equation essentially integrates the reflected light beam with cross-sectional radius $\omega$ at a distance of the projection plane, over a circular cross section $s_{area}$ of the phototransistor.

$$P = P_E \frac{S_{area}}{\pi \omega^2} \quad (4)$$

where,

$$\omega = f(d)$$

$$\begin{bmatrix} k_1 & k_2 \\ k_3 & k_4 \end{bmatrix} \begin{bmatrix} i_1 \\ i_2 \end{bmatrix} = \begin{bmatrix} \alpha \\ \gamma \end{bmatrix} \quad (5)$$

In order to conceptualize whether light intensity values from two sensors can be used to uniquely estimate pitch and roll of one rotational unit, a theoretical simulation can be demonstrated using the mathematical model. A series of theoretical pitch and roll values were chosen to measure two sets of values 'd' for each sensor in a rotational unit. From these values, two sets of theoretical intensity values were calculated for the sensors using Eq (4). Next, MATLAB (R2022) was used to generate a calibration matrix by calculating coefficients through a linear regression algorithm, based on a least squares approach. Eq (5) shows the calibration matrix for estimating orientation (pitch ($\gamma$) and roll ($\alpha$)) from the theoretical intensities ($i_1$ and $i_2$) using the coefficients $k_{1-4}$ using the known calibration coefficients, another set of pitch and roll values were used to get intensity values for two sensors (Equation 4).

These were then multiplied by the coefficient matrix to get estimate pitch and roll values ($\alpha$ and $\gamma$). These were plotted against the actual pitch and roll values used, as can be seen in Figure 5. Here it can be seen that the estimation compared to the actual values have substantial overlap. Any deviation that occurs is likely due to not describing an exact model due to the simplifications. Despite this, it can be seen that orientation motion patterns can be estimated with some accuracy using two theoretical intensity values belonging to two sensors. Therefore, as intensity is proportional to voltage induced in the phototransistor of the optoelectronic sensor, it can be deduced that the sensing principle could work in practice and that two sensor voltage values can be used to estimate two orientations during motion.

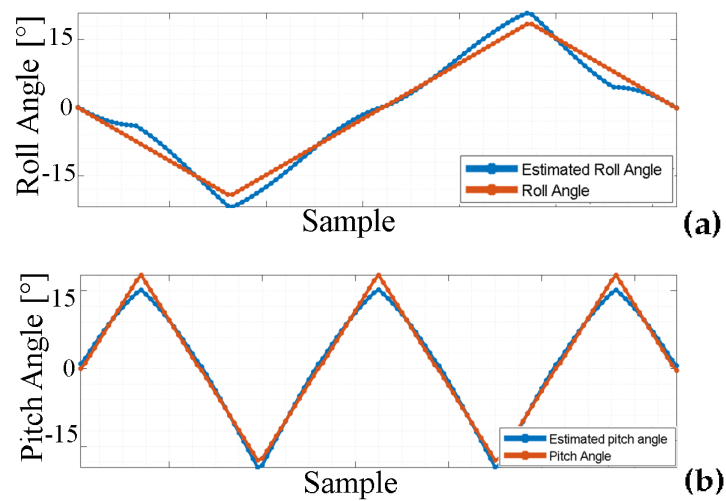

**Fig.5.** Mathematically calculated pitch/roll (blue) compared to simulated motion pitch/roll orientations (orange).

## 4. Experimental Platform

### 4.1 Sensor Calibration Platform

The experimental platform was designed as shown in Figure 2, comprising a section of a flexible tendon actuated robot. For calibration of each pair of sensors at each unit, the process design shown in Figure 6 was used. This utilised 3D printed rigid fixtures that could be screwed around the robotic structure in a way that limits its motion. For example, to calibrate the two sensors on the top unit, the fixtures were fixed onto the three lower units, removing any capability for motion in these units while allowing motion due to tendon actuation in the top unit. A motion pattern was generated to cover all angles in both pitch and roll direction in increments of 0.1° in these two orientations up to ±15°. For example, the pitch is incremented by 0.1°, and motion of the unit in the roll orientation between ±15° is set in increments of 0.1°. The pitch increments again by 0.1°, and the next set of roll orientations are set in motion. This is repeated until all pitch and roll motions are achieved for ±15° each for that unit. During the motion, both optoelectronic sensor voltage readings and IMU orientation data readings were recorded. This provided a full set of sensor readings for the full range of motion for that rotational unit. A set of this data for one rotational unit can be seen in Figure 7. The calibration was carried out by using the IMU data as ground truth, and along with the voltage sensor readings, input into a non-linear regression algorithm to find coefficients that transformed the two sets of sensor voltages to an estimation of pitch and roll values ($\alpha$ and $\gamma$). This data was fed into MATLAB to generate a set of 8 coefficients per orientation estimation, ($k_{1-8}$) and ($j_{1-8}$), to map the two voltage values (*v1*, *v2*) to an estimate of pitch ($\gamma^i$) and roll ($\alpha^i$). This process is repeated for all four units (*i=1 to i=4*), involving each pair of sensors. The equation below (Eq. 6) shows the calculation for the orientation estimations using the sensor voltage data and calibration coefficients. As such, an estimation of pitch and roll could be made at each rotational unit, and therefore, an estimation of the total orientation of the full manipulator could be made. Validation of the calibration results is carried out by setting the full robot segment to move to its maximum angle range of ±60° in both orientations. All sensor voltages are recorded during this motion and multiplied by the calibration coefficients to give estimated orientations during motion which are compared to the orientations given by the IMU sensor. These results are discussed in the following section. Regarding the light intensity model described in section 3, we can further evaluate the validity of this by comparing sensor output voltages during motion to estimated light intensities given the same motion orientation pattern. This is shown in Figure 8, while calibration validation is shown in Figure 9 and Table I.

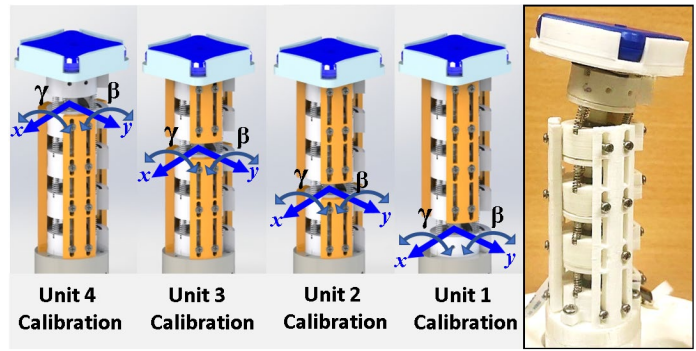

**Fig.6.** Calibration Platform using motion locking fixtures (orange) to allow independent calibration of each unit's sensors during a full range of motion in two orientations. IMU sensor (blue) is attached to the tip unit.

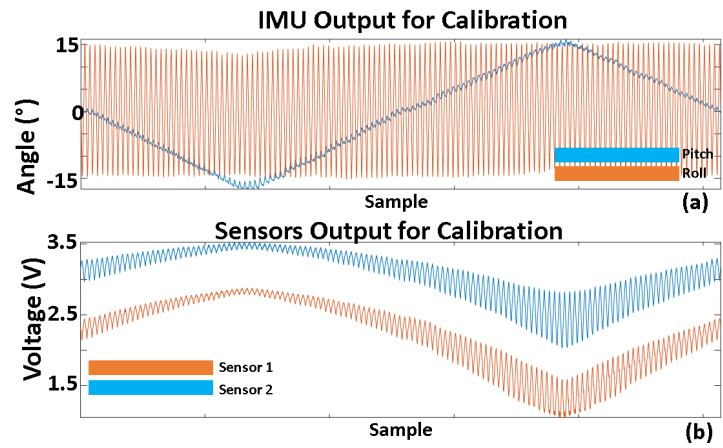

**Fig.7.** Output for single unit calibration, displaying sensor output along with full set of recorded IMU orientation data.

$$\gamma_i = k_{1_i}v_{1_i} + k_{2_i}v_{2_i} + k_{3_i}v_{1_i}^2 + k_{4_i}v_{2_i}^2 + k_{5_i}v_{1_i}v_{2_i} + k_{6_i}v_{1_i}v_{2_i}^2 + k_{7_i}v_{2_i}v_{1_i}^2 + k_{8_i}$$
$$\alpha_i = j_{1_i}v_{1_i} + j_{2_i}v_{2_i} + j_{3_i}v_{1_i}^2 + j_{4_i}v_{2_i}^2 + j_{5_i}v_{1_i}v_{2_i} + j_{6_i}v_{1_i}v_{2_i}^2 + j_{7_i}v_{2_i}v_{1_i}^2 + j_{8_i}$$
$$i = 1,2,3,4 \qquad (6)$$

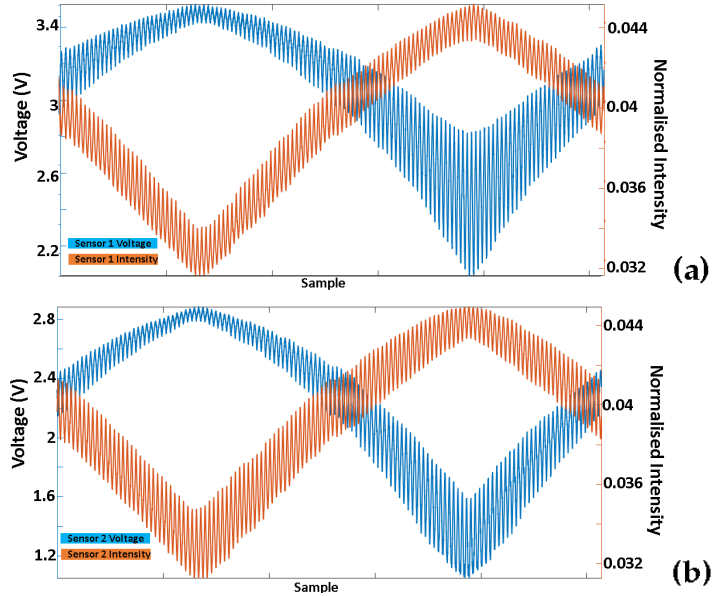

**Fig. 8.** Comparing sensor voltage output with theoretically estimated light intensity for both sensors in one rotational unit of the robotic manipulator.

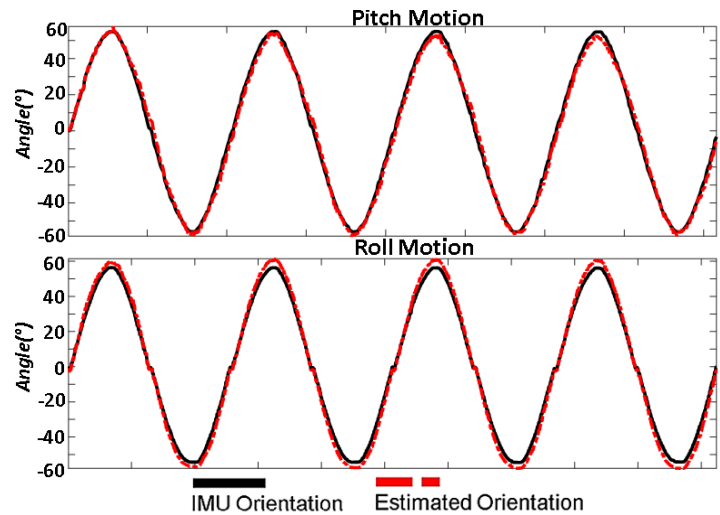

**Fig. 9.** Calibration validation results. Robotic manipulator was moved freely between ±60° in both orientations, and estimated orientation using calibration coefficients is compared to IMU orientation.

TABLE I: CALIBRATION VALIDATION ERROR

| Orientation | % Error | RMS Tip Error (°) | Maximum Tip Error (°) |
| --- | --- | --- | --- |
| Pitch | 0.77 | 2.45 | 6.03 |
| Roll | 0.21 | 3.27 | 6.86 |

## 5. Results and Analysis

Figure 7 shows the output of calibration for one unit of the robotic structure. It shows the recorded IMU data (Figure 7a), with both pitch and roll measurements between ±15° in incremental steps as described previously. This data is recorded along with the two sets of sensor voltages during motion, shown in Figure 7b. As can be seen, the motion data is more stable compared to previously referenced technique [14], that used a PLA printed ball joint as opposed to a spherical bearing as is done here. The use of this bearing eliminated some of the noise previously seen in the sensor data, owing to friction, and the motion is a lot smoother. Next, Figure 8 shows the graphs for both sensors one and two during one rotational unit calibrating motion and compares the sensor voltage output to the estimated relative reflected light intensity. As can be seen, the estimated light intensity generally follows the sensor voltage pattern. It is however inverted relative to the sensor voltage. This is because when considering the interaction of light with the sensor - the intensity collected by the phototransistor is proportional to the current. This current induces a drop in the voltage output of the collector in the phototransistor which corresponds to the voltage values shown in the graphs in Figure 8. This current is inversely proportional to the voltage drop displayed in the graph. This was also seen when comparing Figures 2d and 2e, which both show the characteristic optoelectronic sensor behavior over proximity to a reflective surface, where Figure 2d displays the output voltage, and Figure 2e displays the inverse of this, as the relative output current. As such, we must update the theoretical model to find a relation between the sensor voltage output and the theoretical intensity currently described by the model.

In another aspect of this comparison (Figure 8), we can see that sensor 1 and 2 voltage outputs differ in range and starting voltage level, whereas for the estimated theoretical intensity output, the starting values are almost identical in range and starting value- indicating symmetry between the two sensors' output during motion and corresponds to what is described by the mathematical model. A reason for which this is not reflected in the real sensor output may be due to non-symmetric placement of the optoelectronic sensor pairs on the PCB strip (Figure 2g), in regard to the orientation of the LED and Phototransistor. This resulted in slightly different sensor responses, which may have affected the sensing system properties in terms of range and sensitivity. This can be rectified in future prototypes through reorienting the optoelectronic sensors on one side of the flexible PCB strips. This could, in the future, show a more predictable sensor voltage output when comparing it to the theoretical model. Nonetheless, it can be shown that the model may predict tendencies displayed practically in the experiments.

For the calibration validation results shown in Figure 9, it can be seen that the orientation estimation was successful. The maximum motion range of ±60° in purely pitch, as well as roll orientations were carried out over four cycles. The orientation estimations based on the calibration coefficients calculated through a non-linear regression model (Eq. 6) closely follow the orientation given by the IMU sensor, with maximum RMS error of 3.27°, and percentage error of 0.77% (Table I). The maximum tip orientation estimation errors for the pitch motion shown in Figure 9 was 6.03°, and 6.86° for the roll motion. As a measure of repeatability, the mean standard deviation for pitch orientation was 1.3128° and 1.0274° for roll, over four cycles of motion between ±60°. The estimation of the shape error is relatively successful, however small errors along the points summate to larger tip positional and orientation estimation errors. It is possible that these errors at each point may arise from interference between the sensor pairs. This interference refers to a case where infrared light emitted from the LED of one optoelectronic sensor may be reflected into the phototransistor of the adjacent optoelectronic sensor. This can affect calibration and orientation estimation accuracy. As such, future prototypes will address this issue by designing a new PCB strip with a current switching circuit using a demultiplexer, to eliminate this interference effect. This would involve alternating power between adjacent optoelectronic sensors at a high frequency, meaning that adjacent sensors are never on at the same instance. This would also have the advantage of reducing overall power consumption, which supplements the safety of the system for clinical application.

## 6. Conclusion and Future Works

To conclude, it can be said that the optoelectronic based shape sensing technique coupled with a spherically shaped reflector is successful in orientation estimation in two degrees of orientation. The sensing configuration in this system is greatly reduced in size and utilises fewer sensors while successfully estimating orientation in the same motion range of the robotic manipulator, compared to previous prototypes. The use of a single flexible circuit eliminates wiring for a more simplified design, while use of a spherical bearing for the robot structure allows for smoother motion and subsequently better calibration results. For future improvement in consequent prototypes, certain features can be addressed to improve suitability to targeted application. In this prototype, the size was left to be larger in order to easily test different parameters such as the sensor placement, and in carrying out of the mechanical calibration process. However, once this shape sensing technique is established and proof of concept has been achieved and the principles of the shape sensing are understood, the future designs would of course be scaled down, and its safety pertaining to large-scale electric current from multiple optoelectronic sensors in all segments should be guaranteed. Regarding the size issue, the robot structure can feasibly be miniaturized through the use of smaller spherical bearings, to a diameter of 12 mm. While most MIS based continuum robots require smaller diameter, some applications have utilised robots of up to 12 mm [17][18], while applications such as airplane engine inspection and repairs, as well as for exploration, have used continuum robots with diameters between 12 – 25mm [19-22]. Regarding the safety issue, as well as the interference issue, use of a current source switching circuit using a DEMUX and transistors is proposed to provide all the current required for all the optoelectronic sensors, thereby reducing overall power consumption in the robotic manipulator, and eliminate possibility of interference between adjacent optoelectronic sensors. The maximum angle measurement range of this shape sensing mechanism is limited due to a lower number of units in a one segment (currently 4), so more units will be added (2) to cover a larger measurement range (up to 90°). A soft cover will be used to enclose the continuum robot, to protect it further from any external light, as well as from any damage from chemicals or substances, and to also maintain accurate calibration. Temperature compensation circuits will be implemented to the optoelectronic sensors when future experiments are carried out with more realistic clinical conditions. Nonetheless, the results are promising and show potential for further development of shape sensing in flexible robotic applications. The channel shape experiments enabled us to verify shape sensing performance and showed that the shape sensor is able to detect different curves and is able to respond to shape change due to external forces or obstruction. For testing in more realistic conditions, a full-length robot will be designed integrating this shape sensor for testing in a clinical phantom environment.